\documentclass[a4paper,11pt]{article}
\usepackage{pos}
\usepackage[lofdepth,lotdepth]{subfig}
\newcommand{\marty}{\emph{MARTY}}

\title{\marty, a new C++ framework for automated symbolic calculations in Beyond the Standard Model physics}
\ShortTitle{\marty}

\author*[a]{Grégoire Uhlrich}
\author[a,b,c]{Farvah Mahmoudi}
\author[a,b,c]{Alexandre Arbey}

\affiliation[a]{Universit\'e de Lyon, Universit\'e Claude Bernard Lyon 1, CNRS/IN2P3, \\
Institut de Physique des 2 Infinis de Lyon, UMR 5822, F-69622, Villeurbanne, France}

\affiliation[b]{Theoretical Physics Department, CERN, CH-1211 Geneva 23, Switzerland}
\affiliation[c]{Institut Universitaire de France, 103 boulevard Saint-Michel, 75005 Paris, France}

\emailAdd{g.uhlrich@ipnl.in2p3.fr}

\abstract{Theoretical calculations Beyond the Standard Model (BSM) constitute a challenge for high energy physicists, but are necessary when searching for New Physics. The predictions of a BSM scenario need to be compared with experimental data and the Standard Model values in order to identify the model that fits better what we observe in particle colliders. BSM predictions require very involved and error prone calculations of amplitudes, cross-sections and Wilson coefficients. Calculations at the one-loop level are often necessary for these quantities since some phenomenologically important processes may not occur at tree-level, such as Flavor Changing Neutral Currents (FCNC) in flavor physics that vanish at the tree-level. One-loop calculations have to be done analytically which is very time consuming and in practice rarely done for general BSM models. 
Here we present \marty, a public and open-source C++ code. \marty\ is the very first independent program able to calculate automatically amplitudes, squared amplitudes and Wilson coefficients at the one-loop level for general BSM models. This type of calculations requires a computer algebra system and could only be done, up to now, using Mathematica, a commercial and closed software for symbolic manipulations. \marty\ does not rely on Mathematica since it re-implements its own computer algebra system also in C++, called \emph{CSL}. \marty\ will considerably ease BSM studies as by automating the analytical calculations the main task of the user would be the model building part. Once interfaced with other phenomenological codes in particle physics, \marty\ will be incredibly efficient to make detailed predictions for general BSM models automatically.
}

\FullConference{%
  40th International Conference on High Energy physics - ICHEP2020\\
  July 28 - August 6, 2020\\
  Prague, Czech Republic (virtual meeting)
}

\begin{document}
\maketitle

\section{Introduction}

Calculations in physics Beyond the Standard Model (BSM) have always been a challenge, especially at the one-loop order. Transition amplitudes, differential cross-sections, Wilson coefficients, etc. need to be evaluated in order to make predictions for physical observables. Then statistical comparisons of the predictions against the SM and experimental values allow us to discriminate between BSM models and constrain their parameters.

Calculation at the one-loop level is often necessary, as some physically interesting processes only appear at this order. This is the case of Flavor Changing Neutral Currents (FCNC) in flavor physics, that generally require one-loop level values for Wilson coefficients. At the one-loop level, automatic calculations cannot be performed numerically, and the very large number of terms together with renormalization processes require symbolic computations, as one would do by hand.

Up to now -- at the one-loop level -- symbolic calculations could be done mainly by hand or using Mathematica \cite{mathematica}, a commercial and closed software dedicated to this task. Several packages written with Mathematica have implemented high-energy physics features. FeynRules \cite{feynrules} computes Feynman rules from a BSM Lagrangian, from which FormCalc \cite{formcalc} can derive tree-level and one-loop quantities such as transition amplitudes or differential cross-sections. Packages such as FormFlavor \cite{formflavor} or FlavorKit \cite{sarah,flavorkit} can also calculate Wilson coefficients using the FormCalc machinery.

Many efforts have been made to develop independent computer algebra systems for high-energy physics. LanHEP \cite{lanhep}, CompHEP \cite{comphep} and CalcHEP \cite{calchep} automate together calculations from the Lagrangian using their own symbolic computation framework. However Mathematica packages are still the only ones to provide one-loop calculations and Wilson coefficients in BSM scenarios.

With \marty\ \cite{marty} we introduce a modern solution to this issue. Having for the first time a unique, free and open-source code fully written in modern C++ (2017 standard) implementing all the theoretical BSM machinery up to the one-loop order including Wilson coefficients and Next to the Leading Order (NLO) calculations, and without depending on any other framework, since \marty\ has its own C++ Symbolic computation Library (\emph{CSL}) to manipulate mathematical expressions. 

\section{Design}

\begin{figure}[h!]
    \centering
    \includegraphics[width=0.75\linewidth]{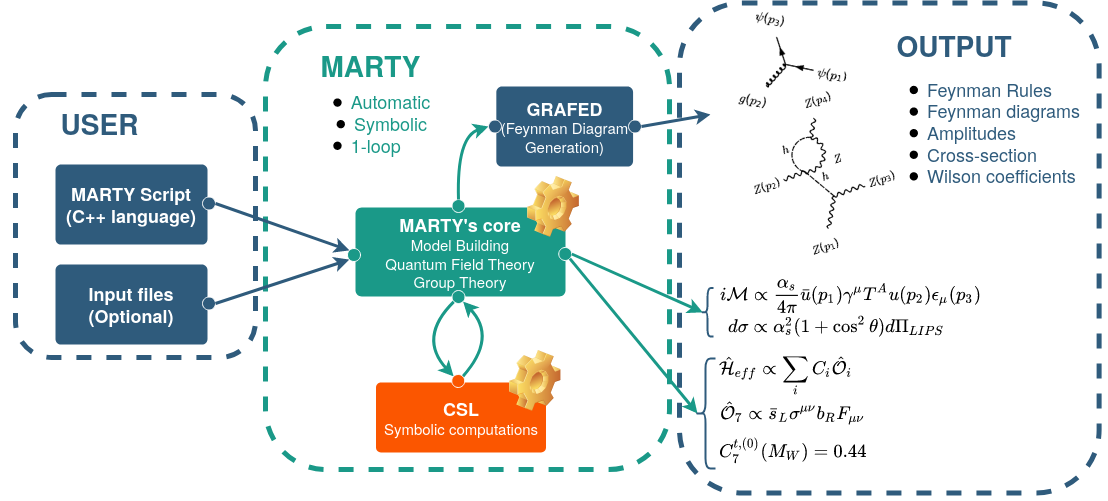}
    \caption{General design of \marty.}
    \label{fig:marty}
\end{figure}

\marty\ is composed of three modules (see Figure \ref{fig:marty}). \emph{CSL} is a pure symbolic calculation library, containing no physics. It includes features allowing to store and manipulate mathematical expressions, in particular with indicial tensors (such as $\gamma-$matrices $\gamma ^\mu _{\alpha\beta}$). Then the physics core of \marty\ implements model building utilities and all the group theory and quantum field theory features needed to perform one-loop calculations for BSM physics. Finally, \emph{GRAFED} (Generating and Rendering Application for FEynman Diagrams) generates automatically Feynman diagram layouts, and displays them via a C++/Qt \cite{qt} application\footnote{\emph{GRAFED} uses only the free and open-source part of Qt.}.

Every calculation in \marty\ is automatic, symbolic, and at the one-loop level. Possible outputs are Feynman rules (from a Lagrangian), transition amplitudes, differential cross-sections, and Wilson coefficients. Inputs have to be given in a \emph{MARTY script} written in C++ language.

\section{Features}
\subsection{Model building in BSM}

\paragraph{BSM scenarios in \marty}
\marty's models are set in a 4-dimensional space-time. Particles of spin 0 (scalar bosons), 1/2 (Weyl, Dirac, or Majorana fermions), and 1 (vector bosons) can be defined. In terms of group theory, the gauge may be any combination of semi-simple groups\footnote{Semi-simple groups are defined as groups with a semi-simple algebra.}: $SU(N), SO(N), Sp(N), E_6, E_7, E_8, F_4, G_2.$ A particle in \marty\ may be any irreducible representation of the chosen gauge, defined by Dynkin labels \cite{dynkin}. This capability allows the user to study a large variety of BSM scenarios.

\paragraph{Model building utilities}

There are multiple methods to give \marty\ information about a BSM scenario. The two most straightforward ways are to give, as for the existing Mathematica packages \cite{feynrules, formcalc, sarah}, either the full Lagrangian, or the explicit list of Feynman rules. Once the Lagrangian is defined, \marty\ will automatically derive Feynman rules.

There is another possibility for models with many terms in the Lagrangian, which is typically the case in many BSM scenarios. Giving the explicit list of interaction terms (or Feynman rules) can be rather long and fastidious. In order to have all factors and signs correct it is often necessary to spend a significant amount of time reading different articles that possibly use different conventions. To overcome this problem, we can use the fact that high-energy Lagrangians are much simpler than low-energy ones: Symmetries are conserved and interaction terms are then written in a much more compact way. They are also easier to find in the literature. It is possible within \marty\ to define such Lagrangians, and use its built-in functions to make it re-derive the final (low-energy) Lagrangian. These functions include flavor- and gauge-symmetry breaking, field replacement / rotation, mass matrix diagonalization, and more.

\subsection{Calculations in BSM}

All calculations performed by \marty\ are automatic and symbolic. It means in particular that no approximation is made, and that the user does not need to perform any task beside providing inputs. Amplitudes, differential cross-sections, and Wilson coefficients in effective theories can be calculated. These quantities may be fully simplified automatically at tree-level, or at one-loop with at most five external particles. One-loop momentum integrals, once reduced symbolically to scalar quantities, are evaluated numerically using the dedicated Fortran library LoopTools \cite{looptools}.

For the first release of \marty, only Leading Order (LO) contributions are included. One-loop results from \marty\ can be used when the tree-level quantities vanish, such as FCNC processes in the SM. One-loop corrections to tree-level processes will be treated in a next version. These will include renormalization (couplings, masses, fields), Renormalization Group Equations (RGEs) and operator mixings for Wilson coefficients.

\marty\ generates C++ code automatically, allowing the user to perform phenomenological analyses. Any mathematical expression (\marty's output typically) may be written as numerical functions in a C++ library, taking as parameter every symbol that has no specified value. These libraries are also built automatically and may be called from a C++ program or directly using Python to facilitate further analysis.

\section{Validation}

\subsection{Cross-section at the tree-level}

\begin{figure}[h]
    \centering
    \subfloat[Diagrams for $ee\to\mu\mu$.]{
        \raisebox{1.2cm}{\includegraphics[width=0.25\linewidth]{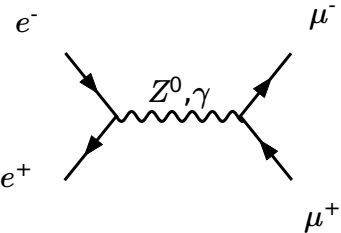}}
        \label{fig:ee_to_mumu_diag}
    }
    \hspace{0.5cm}
    \subfloat[Cross-sections obtained with \marty.] {
        \includegraphics[width=0.6\linewidth]{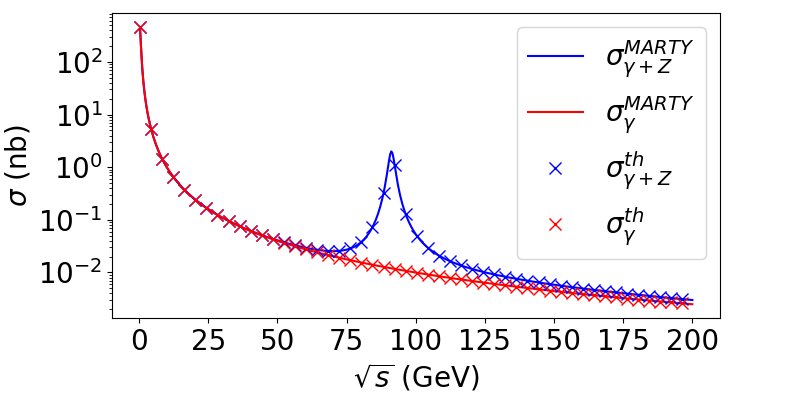}
        \label{fig:ee_to_mumu}
    }
    \caption{$\sigma ^{\text{SM}}\left(e^+e^-\to\mu^+\mu^-\right)$ at tree-level. Results with (blue) and without (red) the $Z$ boson for \marty's output (solid lines) and analytical values (crosses) match perfectly. The mean relative error is $\left\langle\Delta\right\rangle\approx 10^{-6}.$ The $Z$ resonance can be seen around $\sqrt{s} = M_Z = 91.2\ \text{GeV}.$}
\end{figure}

Figure \ref{fig:ee_to_mumu_diag} shows the two diagrams contributing to the $e^+e^{-}\to\mu^+\mu^-$ process at tree-level in the SM. They are both s-channels, with a photon or $Z-$boson exchange. The cross-section is thus shaped around two resonances: The photon resonance ($m_\gamma = 0\text{ GeV}$) and the $Z$ one ($M_Z \approx 91.2\text{ GeV})$. The full cross-section calculation has been done with \marty. The bare output is the differential cross-section. The phase space integration is not done automatically but is straightforward to implement in Python from \marty's output, a numerical function taking external momenta as parameters. Results are compared to analytical formulas calculated by hand, with or without the $Z$ contribution. This is shown in Figure \ref{fig:ee_to_mumu}.

The $Z$-resonance appears clearly, with a maximum value $\sigma _{peak} \approx 2\text{ nb}$. The symbolic part of the calculation (from the Lagrangian to the squared amplitude) is performed automatically by \marty. From the squared amplitude to the final cross-section value, purely numerical computations are performed and no theoretical subtleties are left. This part is deferred to the user and can be done either in C++ or in Python using libraries generated by \marty.

\subsection{Wilson coefficient at the one-loop level}

A more complicated example is the FCNC $b\to sg$ process. We consider here non-primed coefficients only\footnote{Primed coefficients are suppressed by $m_s/m_b$ and are often approximated to zero (when $m_s$ is set to zero).}.
\begin{figure}[h!]
    \centering
    {
        \includegraphics[width=0.19\linewidth]{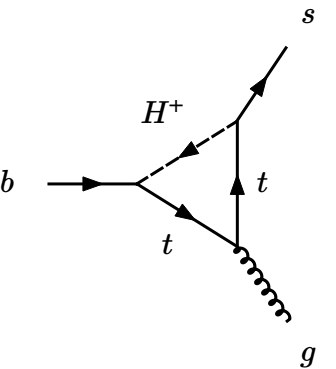}
    }
    \hspace{1cm}
    {
        \includegraphics[width=0.19\linewidth]{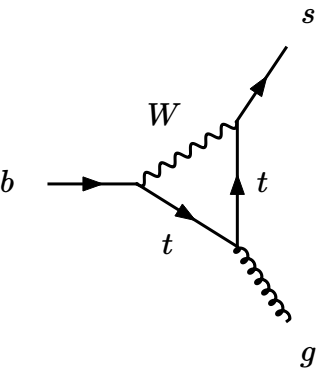}
    }
    \hspace{1cm}
    {
        \includegraphics[width=0.19\linewidth]{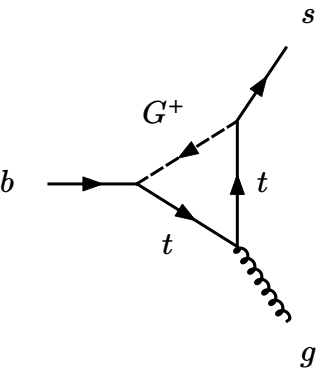}
    }
    \caption{Examples of contributions to the Wilson coefficients $C_8$ in the 2HDM. Diagrams are generated with \emph{GRAFED}. $G^+$ is the Goldstone boson of $W$, and there are in addition charged Higgs $H^+$ contributions specific to the 2HDM model.}
    \label{fig:operator_diags}
\end{figure}

The Leading Order (LO) is at the one-loop level and we perform the calculation on-shell keeping 1-Particle Reducible (1PR) diagrams. Calculation is done with \marty, and the results are compared to the analytical formulas taken from the Superiso manual \cite{superiso}.

\begin{figure}[h!]
    \centering
    \includegraphics[width=0.78\linewidth]{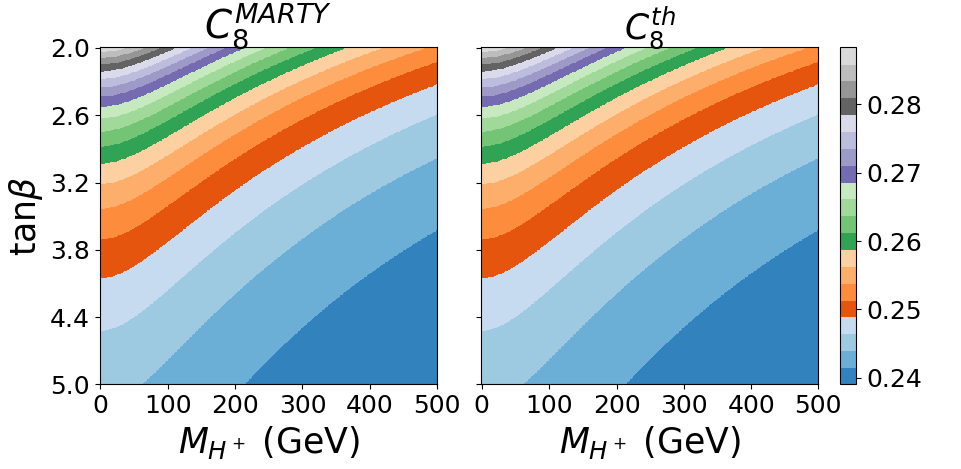}
    \caption{$C_8^{2HDM}(M_W)$ as a function of the charged Higgs mass $M_{H^+}$ and $\tan\beta$ for \marty\ on the left and analytical values on the right \cite{superiso}. The mean relative error is $\left\langle\Delta\right\rangle\approx 10^{-5}.$}
    \label{fig:c72hdm}
\end{figure}

We performed this calculation with \marty\ in the 2 Higgs Doublet Model (2HDM) of type IV (Lepton specific). There are SM ($W$ and Goldstone $G^+$) and charged Higgs ($H^+$) contributions which depend on the charged Higgs mass $M_{H^+}$ and $\tan \beta$ that is the angle between the vacuum expectation values of the two Higgs doublets. Examples of contributions are presented in Figure \ref{fig:operator_diags}. We vary $M_{H^+}$ in $[0, 500]\text{ GeV}$ and $\tan\beta$ in $[2, 5]$. Results are presented in Figure \ref{fig:c72hdm}, where it can be seen that \marty's output is in perfect agreement with analytical formulas.

\section{Conclusion and Outlook}

We presented \marty, a new C++ framework automating theoretical calculations symbolically in BSM physics. The degree of generality reached by \marty\ has never been obtained before. It has its own symbolic manipulation library (\emph{CSL}), and automates all theoretical calculations directly from the Lagrangian. Feynman rules, Feynman diagrams, amplitudes, cross-sections, and Wilson coefficients can be obtained in a very large variety of BSM models up to the one-loop level. A full NLO treatment will also be implemented in the near future, what is lacking in currently available codes.

A proof of its capabilities has been demonstrated by a tree-level cross-section calculation in the SM, and a 1-loop Wilson coefficient calculation in the 2HDM. Most of popular BSM models can be built in \marty. The Minimal Supersymmetric extension of the Standard Model (MSSM), extended gauge models, vector-like quarks are examples of possible BSM implementations.

\marty\ can be very useful for BSM phenomenology. One of its most significant advantages is to be written as a unique code, not depending on any other framework. Within \marty\ every aspects of model building and high-energy physics calculation are under control. This is a unique opportunity for more collaborations to take this code even further, extending it to new models, other simplification methods, or even different types of calculations.

\bibliographystyle{h-physrev5}  
\bibliography{references}  

\end{document}